\documentclass[manuscript]{aastex}

\begin{document}


\title{Far-Ultraviolet Number Counts of Field Galaxies\\}


\author{Elysse N. Voyer\altaffilmark{1}}
\affil{$\textit{Department of Physics, The Catholic University of America,
Washington, D.C. 20064}$}
\affil{$\mathtt{48voyer@cardinalmail.cua.edu}$}

 \author{Jonathan P. Gardner}
 \affil{$\textit{Astrophysics Science Division, Observational Cosmology
Laboratory, Goddard Space Flight Center,}$}
 \affil{$\textit{Code 665, Greenbelt, MD 20771}$}

 \author{Harry I. Teplitz}
 \affil{$\textit{Spitzer Science Center, California Institute of Technology, 
220-6, Pasadena, CA 91125}$}

 \author{Brian D. Siana}
 \affil{$\textit{California Institute of Technology, MS 105-24, Pasadena, CA
91125}$}

 \and

 \author{Duilia F. de Mello\altaffilmark{2}}
 \affil{$\textit{Department of Physics, The Catholic University of America,
Washington, D.C. 20064}$}


\altaffiltext{1}{NASA Graduate Student Research Program Fellow}
\altaffiltext{2}{Observational Cosmology Lab, Goddard Space Flight Center}


\begin{abstract}
The far-ultraviolet (FUV) number counts of galaxies constrain the evolution of the star-formation rate density of the universe. We report the FUV number counts computed from FUV imaging of several fields including the Hubble Ultra Deep Field, the Hubble Deep Field North, and small areas within the GOODS-North and -South fields. These data were obtained with the Hubble Space Telescope Solar Blind Channel of the Advance Camera for Surveys. The number counts sample a FUV AB magnitude range from 21-29 and cover a total area of 15.9 arcmin$^{2}$, $\sim$4 times larger than the most recent HST FUV study. Our FUV counts intersect bright FUV GALEX counts at 22.5 mag and they show good agreement with recent semi-analytic models based on dark matter ``merger trees" by \citet{som11}. We show that the number counts are $\sim$35\% lower than in previous HST studies that use smaller areas. The differences between these studies are likely the result of cosmic variance; our new data cover more lines of sight and more area than previous HST FUV studies. The integrated light from field galaxies is found to contribute between 65.9$^{+8}_{-8}$ -- 82.6$^{+12}_{-12}$ photons s$^{-1}$ cm$^{-2}$ sr$^{-1}$ \AA $^{-1}$ to the FUV extragalactic background. These measurements set a lower limit for the total FUV background light.
\end{abstract}








\section{Introduction}
Measuring the number counts of field galaxies within an observed area as a function of magnitude is one of the fundamental techniques used to study galaxy evolution throughout cosmic time. Galaxy number counts are used to test theoretical models of galaxy evolution; changes in the slope of number count distributions reflect physical changes in the underlying galaxy populations. Such models can predict galaxy properties in various bandpasses and for various redshifts (z). At observed far-ultraviolet (FUV) wavelengths, galaxy counts probe light from unobscured star-formation for z $<$ 1, after which the Lyman limit (912\AA) shifts into the observed bandpass. Little to no UV light is detectable blueward of this limit because it is used in ionizing HI gas in the interstellar and intergalactic medium between the galaxy and the observer. This has been shown in several studies attempting to constrain the Lyman continuum escape fraction at various redshifts \citep[i.e.,][]{sia10,sia07,bri10,cow09}. The majority of the detected UV light is radiated by hot, massive, O and B stars, that have spectral energy distributions peaking at these wavelengths. Due to their short lifetimes, the UV light from O and B stars traces the star-forming regions within galaxies. For this reason, number counts of UV detected galaxies provide a window into ongoing extragalactic star-formation history. 

FUV number count studies are only possible with space-based observations since the Earth's atmosphere is opaque to UV light. Over the past two decades, only a handful of space-based field galaxy surveys have been carried out at UV wavelengths \citep{mil92, deh94, gar00, igl04, xu05, tep06, hov09} since long integration times are required to reach faint magnitudes. The first UV galaxy counts were measured by \citet{mil92} using the balloon-borne FOCA instrument at 2000\AA\ and bright magnitudes 15--18.5, covering a large area of sky ($\sim$6 deg$^{2}$). Later, two studies used deep imaging from the Hubble Space Telescope (HST) to measure faint UV galaxy counts. \citet{gar00} measured NUV (2365\AA) and FUV (1595\AA) counts over smaller areas (1.54 arcmin$^{2}$) for magnitudes 24.5--29.5 in the Hubble Deep Fields North and South (HDF-N \& -S). \citet{tep06} measured FUV (1600 \AA) counts for magnitudes 20.5--28.5 in the HDF-N, covering 3.77 arcmin$^{2}$. Bright UV galaxy counts (NUV: 2310, FUV: 1530), between 14--23.8 mag, were measured by \citet{xu05} using 36 Medium-depth Survey fields (MIS) and 3 Deep Survey fields (DIS) obtained with the Galaxy Evolution Explorer (GALEX). They cover a total area $\sim$20 deg$^{2}$. More recently, \citet{hov09} used the \textsl{Swift} UV/Optical Telescope to measure NUV (1928\AA, 2246\AA, 2600\AA) galaxy counts in a 289 arcmin$^{2}$ area of the Chandra Deep Field South (CDF-S) between 21--26 mag. However, the only two studies measuring the faint end slope (24.5--29.5 mag) of the FUV galaxy counts are subject to cosmic variance effects, due to the small areas surveyed, and known overdensities in the HDF-N \citep{coh00}.

In this paper we present FUV (1614\AA) galaxy number counts from deep images obtained with HSTs Solar Blind Channel (SBC) on the Advanced Camera for Surveys (ACS). These observations sample a magnitude range of 21.5--29.5 and cover an area $\sim$4 times larger (15.9 arcmin$^{2}$) than the most recent FUV study that previously covered the largest area at these wavelengths and magnitudes \citep{tep06}. In Section \ref{data} we present the data used for this study. In Section \ref{nc} we discuss the measurement of the number counts and corrections to the counts due to observational biases. The number counts are compared with previous studies in Section \ref{compare} and theoretical models in Section \ref{compare2}. Cosmic variance is discussed in Section \ref{cosmicvar}, and the FUV extragalactic background light (EBL) calculation is presented in Section \ref{ebl}. Results of this study are summarized in Section \ref{sum}.

\section{The Data}
\label{data}
\begin{figure*}
\epsscale{1}
\plotone{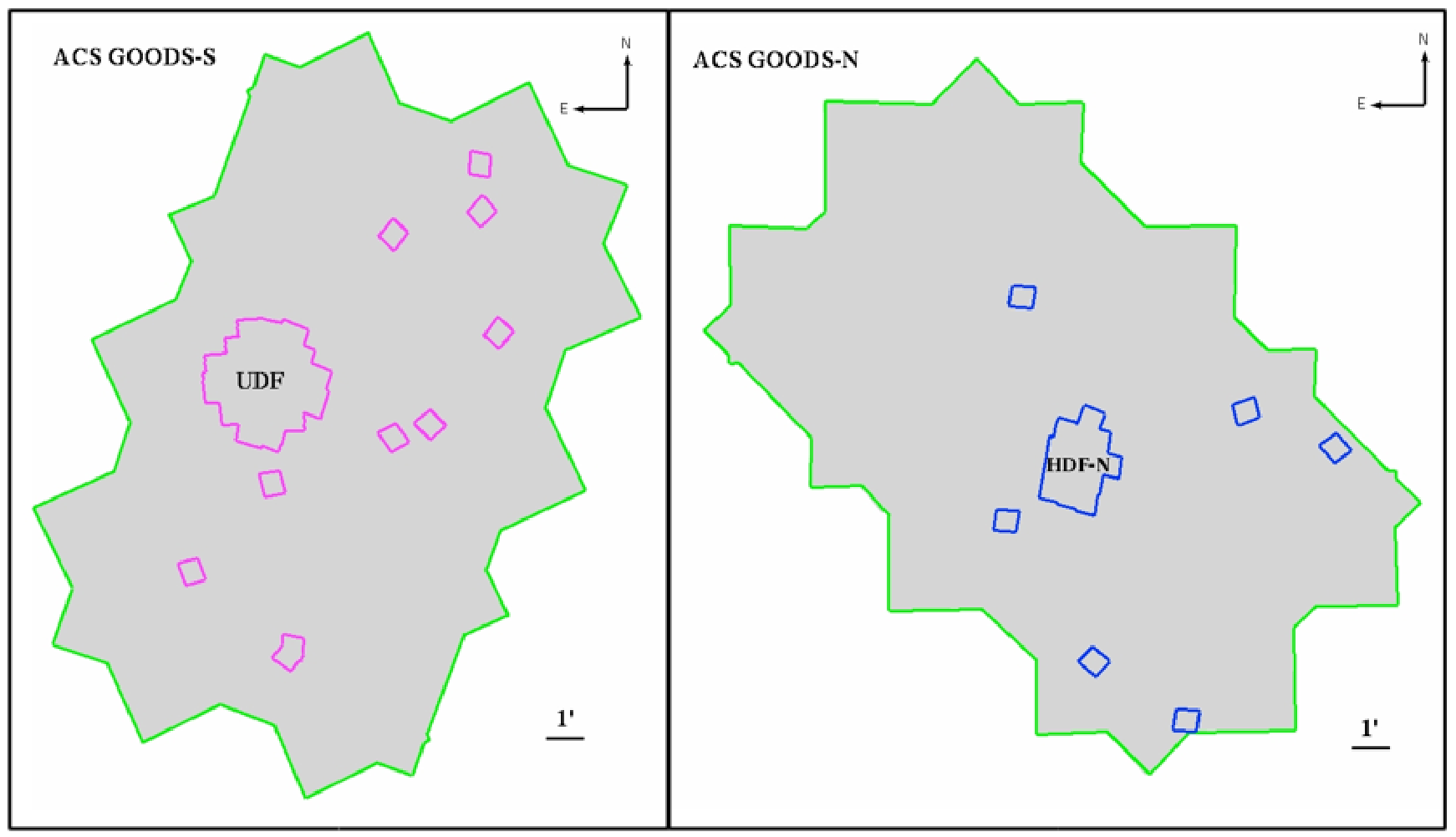}
\caption{Footprints of regions observed with the ACS SBC within the ACS GOODS-N and -S areas.}
\label{foot}
\end{figure*}

For this study we used FUV observations from three different data sets: the HDF-N area  of the Great Observatories Origins Deep Survey North (GOODS-N) field, the Hubble Ultra Deep Field (HUDF) area of the GOODS-South (GOODS-S) field, and smaller fields in various parts of the GOODS-N and -S fields (see Figure \ref{foot}). The HDF-N data is from the HST General Observer Program 9478, the HUDF data is from the HST Cycle 13 Treasury Program 10403, and the smaller GOODS-N and -S fields are from the HST Cycle 15 General Observer Program 10872. All observations were obtained with the SBC detector on Hubble's ACS. The ACS SBC detector is a Multi-Anode Microchannel Array (MAMA) with a field of view of 34$''$.6$\times$30$''$.8. All observations were taken through the long-pass quartz filter, F150LP, that peaks $\sim$1500\AA, has a bandwidth of $\sim$550\AA, effective wavelength of 1614\AA, and a FWHM=177\AA. At z $\sim$ 0.6 the Lyman limit, 912\AA, is bandshifted to 1500\AA, thus the SBC F150LP is only sensitive to the brighter galaxies beyond z $>$ 0.7.

Final images of the HDF-N and HUDF used for source detection were constructed using the DRIZZLE package in IRAF\footnote{IRAF is distributed by the National Optical Astronomy Observatories, which is operated by the Association of Universities for Research in Astronomy, Inc., under cooperative agreement with the National Science Foundation.}. The smaller GOODS-N and -S images were tiled onto the original GOODS areas for source detection. Photometry is performed using similar procedures to those in \citet{gar00a} and \citet{tep06} where optical segmentation maps produced with the SExtractor software package \citep{ber96} are used to determine the pixels that are included in the FUV flux measurement. SExtractor has difficulty working on data with few counts per pixel \citep{gar00a}, such as the FUV images, thus in order to prevent false segmentation of sources, the optical image is used for detection because galaxies morphologies are less clumpy in the rest-frame optical  than in the UV \citep{tep06}. We used SExtractor to detect sources in the GOODS-N \& -S fields ACS F606W (V-band) images and defined extraction isophotes that extend out to where the galaxy flux per pixel is 0.8 times the background RMS ($\sigma$). The 0.8$\sigma$ apertures are then used to extract fluxes in the FUV images. The F606W images \citep{bec06,gia04} are more than one magnitude deeper than the FUV images (in AB mags). Thus, we can be confident that we are capturing all of the FUV flux within these apertures.

Galactic extinction does not vary significantly over the areas we observe because the GOODS fields were selected in part for the low extinction along their sight-lines \citep{bec06, wil96}. From the Galactic dust maps of \citet{shl98}\footnote{Accessed via the NASA/IPAC Infrared Science Archive (IRSA) Galactic Dust Extinction tool.} we find the range of extinction to be small over the GOODS-N \& -S fields, varying between 0.0347 $\le$ A$_{V}$ $\le$ 0.0381 and 0.0236 $\le$ A$_{V}$ $\le$ 0.0298, respectively. From these dust maps we find A$_{V}$ at the central coordinate of each FUV source and calculate the corresponding amount of extinction in the FUV, A$_{1610}$, via the ratio given in \citet{sia10} based on the extinction curve of \citet{car89}: A$_{1610}$/A$_{V}$ =2.55. None of the corrections for Galactic extinction are larger than 10\%. 

\begin{figure*}
\epsscale{1}
\plotone{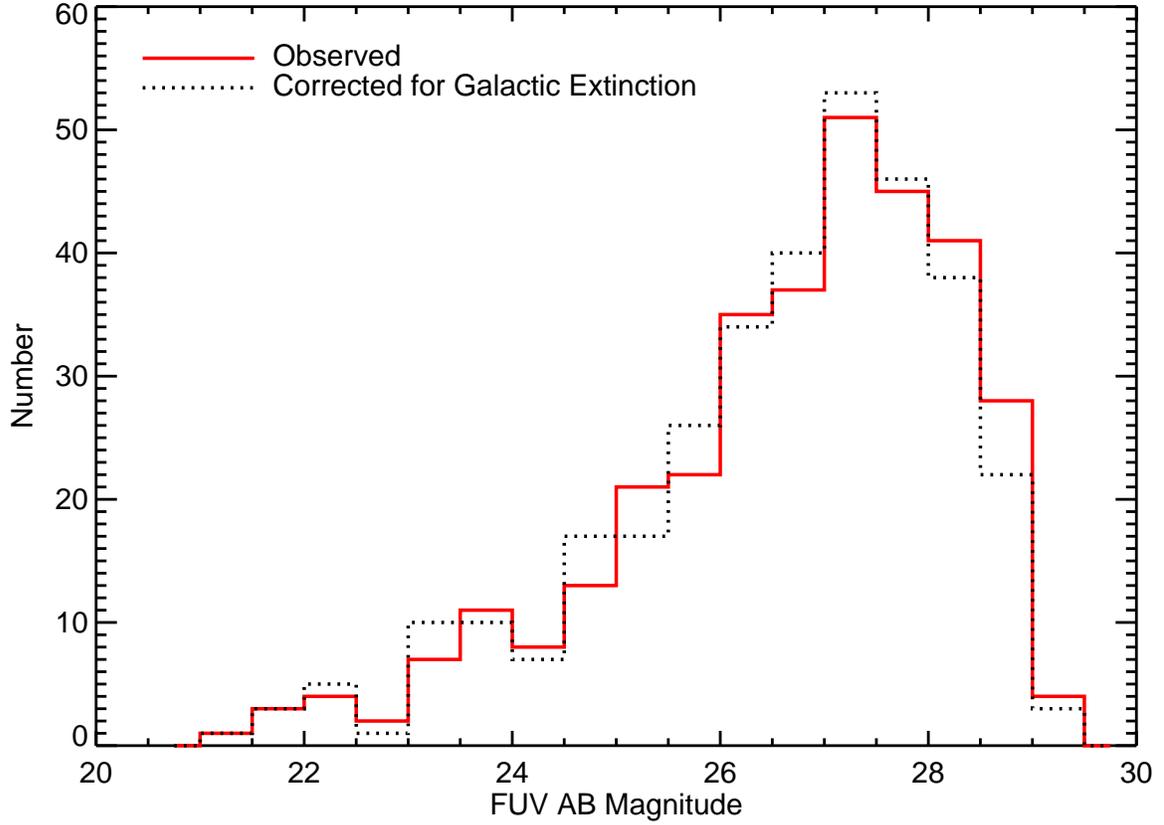}
\caption{FUV magnitude distribution for the 333 sources included in FUV number counts. Both the magnitude distribution as observed and the magnitude distribution corrected for Galactic extinction are shown. The extinction correction was done with A$_{V}$ values from the \citet{shl98} Galactic dust maps and the ratio of A$_{1610}$/A$_{V}$=2.55 calculated in \citet{sia10}.}
\label{magdist}
\end{figure*}

\begin{figure*}
\epsscale{1}
\plotone{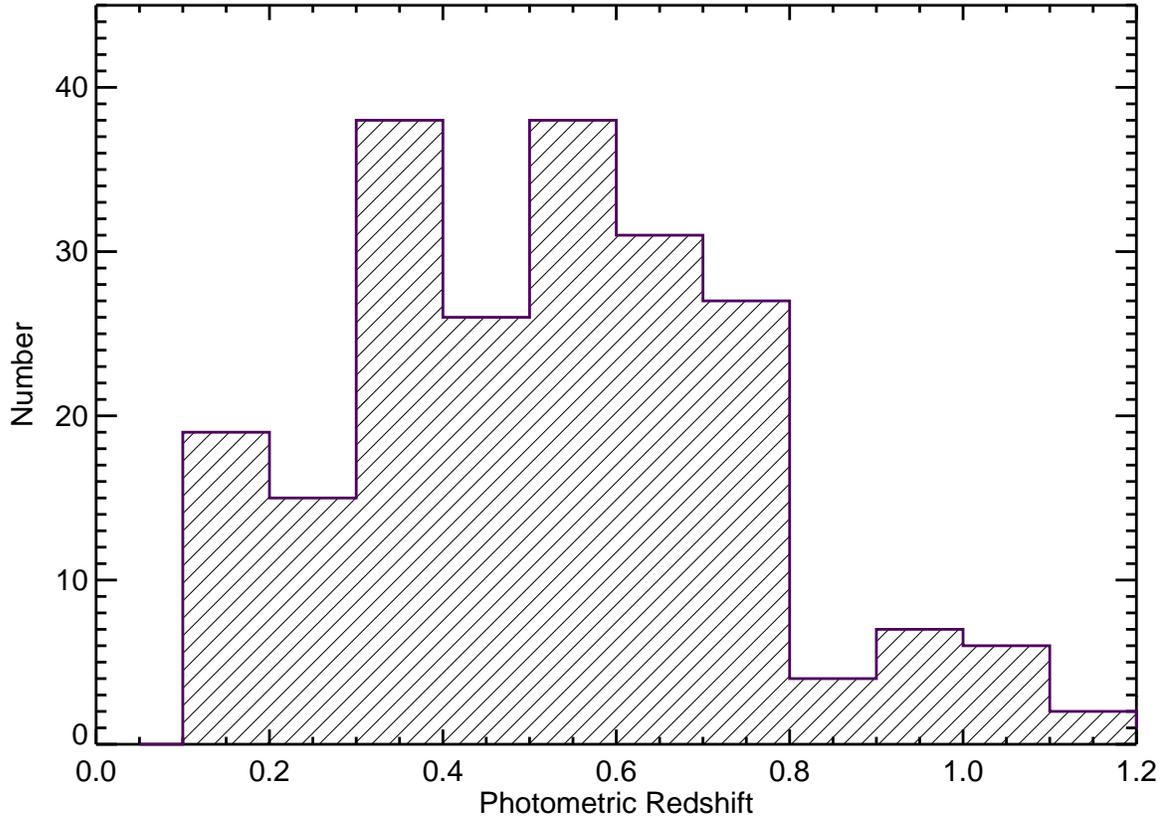}
\caption{Distribution of photometric redshifts, where available, for 212 sources from the FUV number counts sample \citep[][T. Dahlen private communication]{dah10}}
\label{zdist}
\end{figure*}

We detected 114 FUV sources in the HUDF area of GOODS-S, 113 FUV sources in the smaller GOODS-N and -S images, and 116 FUV sources from the HDF-N area of GOODS-N. We removed 10 sources because they were too close to the edges of the images, leaving 333 sources to be included in the measurement of the number counts. Three sources are also $\mathit{Chandra}$ X-ray detections CXO J123648.0+621309, CXO J033239.0-274602, and CXO J333213.2-274241 \citep{eva10} located in the HDF-N area of GOODS-N, the UDF area of GOODS-S, and a smaller area of GOODS-S, respectively. No stars were detected in our sample. The total sample covers an AB magnitude range from 21--29 and its magnitude distribution begins to drop-off at $\sim$28.5 as shown in Figure \ref{magdist}. Redshifts are available for 212 sources \citep[][T. Dahlen private communication]{dah10,coh00} and the distribution for the FUV sample is shown in Figure \ref{zdist}.

\section{Number Counts in the FUV}
\subsection{Measurement of Number Counts}
\label{nc}
In order to measure the number counts of galaxies in our sample we used the method developed in \citet{gar00}. Because there are variations in depth across the FUV images, each FUV source would not necessarily be detectable over an entire image. For each source we must calculate the total area in which it would have been be detected in each image. We use the root-mean-square (RMS) error maps, produced from the weight maps of the drizzled SBC images, to determine these areas. Small-scale-variations created during the image drizzling process are accounted for by smoothing the RMS maps with a 0$''.4\times0''.4$ median filter. Each sources AB magnitude and size are required to calculate the total detection area in an image for that source. FUV AB magnitudes (FUV$_{AB}$) of the sources in all observed fields were obtained from photometric catalogs produced with SExtractor as described in Section \ref{data}. The size of each source was determined from the SExtractor segmentation maps used for the catalog photometry. Using the size and magnitude of each source, we calculate the maximum RMS error of a pixel at 3$\sigma$ in the following way: $\textit{flux}$/(3$\times\sqrt{size}$). Pixels in the RMS map with errors less than or equal to this value make up the total area over which the source would have been be detected at 3$\sigma$.

In order to be consistent, and not overestimate the detection area of the other sources, we cut down the edges of each RMS map by a length equal to the radius of the circular area of each removed source (discussed in Section $\ref{data}$) before calculating their detection areas. This includes both the outer edges of images, as well as edges on the inner parts of the images where drizzled fields do not overlap in the HUDF and HDF-N. In any given magnitude bin there might be a failure to detect low surface brightness objects that are actually there. This effect can become larger towards fainter magnitudes. Thus, number count measurements must be adjusted for such incompleteness biases. To correct for incompleteness we use two independent methods, bootstrap-sampling, and detection of artificially introduced galaxies. For the first method we bootstrap-sample the size distribution of the version 2.0 GOODS-S V-band catalog\footnote{http://archive.stsci.edu/prepds/goods/} starting with a randomly generated FUV galaxy sample. First, 1000 FUV magnitudes are randomly generated for each FUV magnitude bin. Next, following the procedure in \citet{gar00}, we use the mean and standard deviation parameters of a Gaussian distribution fit to the FUV-V color distribution in the HUDF between 24 $\leqslant m_{AB}\leqslant$ 28 to randomly generate FUV-V colors. Even though the completeness correction is ultimately applied to all magnitude bins from 21.5-29.5, this range (24 $\leqslant m_{AB}\leqslant$ 28) is selected for determining the Gaussian distribution because the magnitude distribution of the FUV sample drops off at $\sim$28.5, and there are very few galaxies in the HUDF with magnitudes brighter than 24. Thus, including bins brighter or fainter than these magnitudes would introduce unwanted errors into the distribution. Because the Gaussian color distribution does not vary greatly between the different SBC fields we only sample the HUDF. With these random FUV magnitudes and FUV-V colors we calculate the optical magnitudes of the randomly generated sample, and match them to the closest optical magnitudes of sources in the GOODS-S V-band catalog. The sizes of these objects are then sampled from segmentation maps produced from public GOODS-S V-band images with SExtractor using 0.8$\sigma$ isophotes to define the source areas. These are isophotes within which each pixel in the V-band images is 0.8$\sigma$ above the background noise. We use a 0.8$\sigma$ size isophote because the same is used to define the source areas for the FUV photometry. From the sizes and FUV magnitudes of the random sample we then calculate the maximum RMS pixel error below which each simulated object would be detected, and proceed to calculate the total detection area for each simulated object in the SBC RMS maps. Finally, to get the completeness correction factor for each magnitude bin, we average per magnitude bin the detection areas of the simulated objects (including galaxies with zero detection area), and take the ratio with the average detection area of all real FUV sources in corresponding bins. 

The second incompleteness correction method introduces 500 artificial FUV galaxies into the SBC data for each magnitude bin and recovers them with the same photometry algorithm used for the real data. We simulate these sources using the IRAF task ARTDATA in the NOAO package. We also simulate 500 V-band galaxies using the same software in order to use their isophotal sizes to provide the area in which to measure the flux of the artificial FUV sources. This approach mimics the procedure of the actual FUV photometry. To determine the correct magnitudes of the artificial V-band galaxies, we use the parameters of the same Gaussian FUV-V color distribution discussed above. A FUV artificial source that has S/N $>$ 3.5 is a detection, and the detection ratio equals the number of sources recovered over 500. Finally, to correct for incompleteness the number counts are divided by the detection ratio in each magnitude bin.  

Both methods yield similar incompleteness corrections, within a few percent of one another, in each bin. An average of these two methods is used for the final correction to the number counts.

\subsection{Comparison with Previous FUV Number Counts}
\label{compare}
In Figure \ref{fuvnc} we present the completeness corrected and the raw FUV number counts from this work and past FUV number counts from the literature. Their measured values, errors, completeness, and detection areas per magnitude bin are provided in Table \ref{nctab}. Small number Poisson statistical errors are calculated for each point from \citet{geh86} at the 1$\sigma$ level. The filled circles represent our completeness corrected counts. The open circles represent the raw counts. The upside-down triangles represent counts done with SBC images of the HDF-N from \citet{tep06}. The asterisks represent counts done with HST STIS in the HDF-N and HDF-S from \citet{gar00}. The squares represent counts done with GALEX from \citet{xu05} (hereafter XU05 fields), and the upright triangles represent counts done with GALEX from \citet{ham10} (hereafter HAM10 field). No color corrections are made between the SBC filter which has a central wavelength of 1614\AA\ (filter peak is $\lambda$=1500\AA) and the STIS and GALEX filters that have FUV central wavelengths at 1595\AA\ and 1530\AA\, respectively. As discussed in \citet{tep06}, the color correction between the SBC and GALEX FUV filters would be significant for galaxies at z $>$ 0.50 because the SBC filter is sensitive to a larger volume ($\sim$30\%) than the GALEX filter. This color difference  results in no more than a factor of $\sim$2 ($\sim$ half a magnitude) between the SBC and GALEX number counts. They also discuss that Ly$\alpha$ emitting sources at z $<$ 0.15 could have the opposite effect resulting from the bluer wavelength coverage of the GALEX filter. About 54\% of our FUV sample with z$_{phot}$ are at z$_{phot}$ $>$ 0.50 and $\sim2.3$\% are at z$_{phot}$ $<$ 0.15. The majority (98\%) of sources at z$_{phot}$ $> $ 0.5 are not comparable to GALEX bins because they have fainter magnitudes (FUV$_{AB}$ $>$ 24). Thus, comparisons with GALEX FUV number counts are not largely affected by ignoring the filter color correction. 

Our galaxy sample probes the faint-end of the FUV number counts, with the majority of sources occupying magnitude bins 23.5-28.5. This is reflected in the error bars of these plotted points. The three faintest objects in our sample have FUV$_{AB}$ = 29.19, 29.21, and 29.33. Although these sources are fainter than the magnitude drop-off of the FUV data ($\sim$28.5), they are detected in the GOODS V-band catalog and above the detection threshold of 0.8$\sigma$. Thus, they are included in these number counts. However, due to the few sources detected, the measurement does not accurately represent the number counts at this faint level. On average our number counts are $\sim$35\% and $\sim$36\% lower than the faint HST FUV counts from \citet{gar00} and \citet{tep06}, respectively.  The differences in the measurements are likely the result of cosmic variance which is discussed in  more detail in Section \ref{cosmicvar}.

At the 22.5 magnitude bin the slope of our number counts intersects the faint end of the GALEX HAM10 field counts but not the XU05 field counts, remaining higher than these at all overlapping magnitudes. It is not well understood why the GALEX counts diverge from each other after FUV$_{AB}$ $\sim$21.25, but \citet{ham10} show the divergence can not be due to their source detection/photometry methods, AGN, or cosmic variance between fields. Also, while cluster members in the HAM10 field bias the bright bins of these number counts, they only compose $\sim$2\% of objects in the faintest bin, which represents the limiting depth of the survey. However, massive clusters are known to be associated with many filaments and the number of filaments is directly correlated with cluster mass \citep{pim04}. Thus, \citet{ham10} do not rule out large scale structure behind the massive Coma Cluster as the culprit of their excess galaxy counts.

\subsection{Comparison with Number Counts Models}
\label{compare2}

\begin{figure*}
\epsscale{1}
\plotone{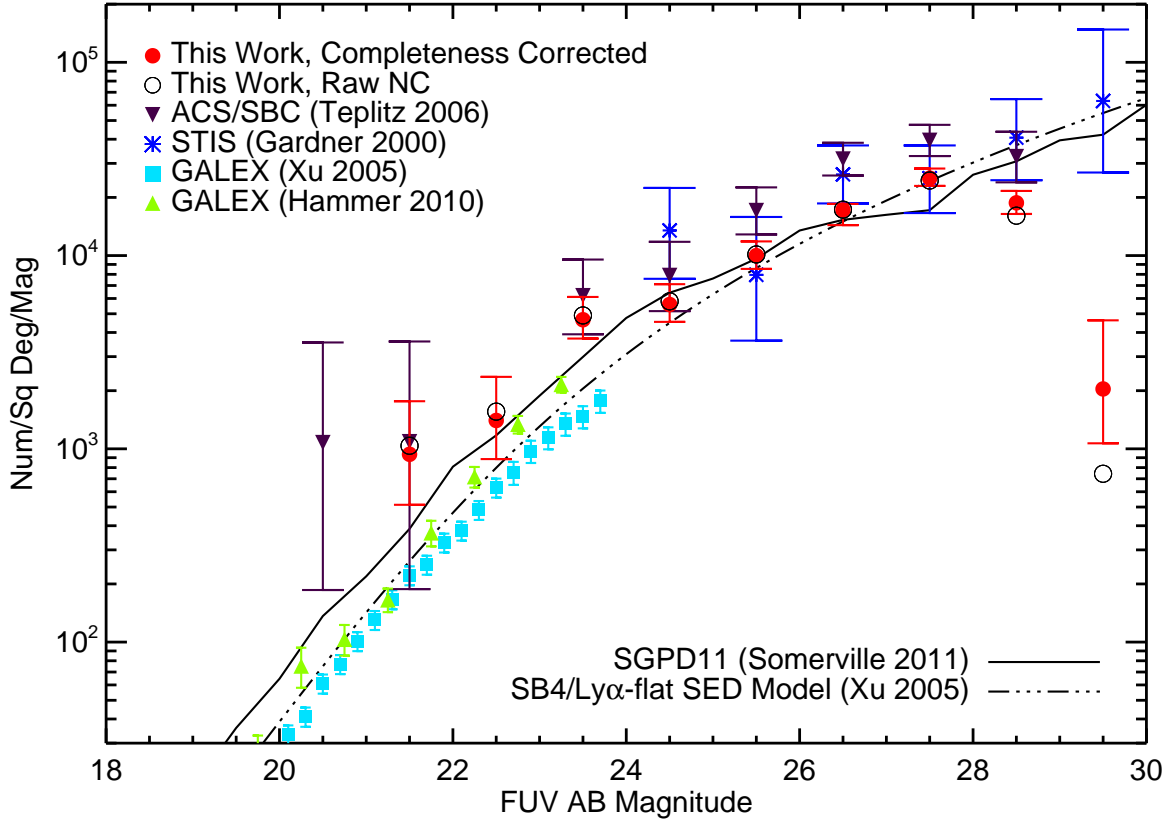}
\caption{FUV number counts of field galaxies from this work shown with FUV number counts from previous studies and compared to semi-analytic models. Error bars are Poisonnian from \citet{geh86}. The caps of the error bars do not reflect an error in magnitude, but have been manually varied in length to better distinguish amongst them.}
\label{fuvnc}
\end{figure*}

A primary use of galaxy number counts is to test and constrain models of galaxy evolution. In Figure \ref{fuvnc} we compare our FUV number counts with two different models, a simple luminosity evolution model from \citet{xu05} and a cosmological semi-analytic model (SAM) from \citet[][hereafter SGPD11]{som11}. The first model is the SB4/Ly$\alpha$-flat SED model. This model is characterized by a UV luminosity evolution, L$^{*}\sim$ (1+ z)$^{2.5}$, and is constructed from a local FUV luminosity function \citep{wyd05} with an estimated K-correction based on the UV SB4 spectral-energy distribution from \citet{kin96} with a flat spectrum between 1200\AA\ and 1000\AA. It was selected as a initial check that our measured number counts were reasonable since this model is in good agreement with evolution models derived from observed luminosity functions at high-z \citep{arn05}. When plotting the SB4/Ly$\alpha$-flat SED model we did not color correct the model from the GALEX FUV effective wavelength at 1530\AA\ to the SBC effective wavelength at 1614\AA\ (see further discussion in Section \ref{compare}).  

The second model, SGPD11, makes use of the latest version of the SAMs developed by Somerville and collaborators \citep{som08,som01,som99}. The backbone of these SAMs are dark matter $``$merger trees" representing the hierarchical build-up of structure in the $\Lambda$ Cold Dark Matter ($\Lambda$CDM) paradigm. The model shown here is the $``$fiducial WMAP5" model presented in SGPD11, and adopts cosmological parameters consistent with the five year WMAP analysis (WMAP5; $\Omega_{m}$=0.2383, $\Omega_{\Lambda}$=0.7617, $\mathit{h}$=0.732, $\sigma_{8}$=0.82). The physical processes included in the model include radiative cooling of gas, photoionization squelching, star formation in quiescent and burst modes, morphological transformation via mergers, supernovae feedback, chemical evolution, black-hole growth, AGN-driven winds, and radio-mode feedback. The UV luminosities for the SAM galaxies are calculated from synthetic SEDs created by convolving the star-formation and chemical enrichment histories for each galaxy with \citet{bru03} stellar population models using a Chabrier initial mass function. A two-component model for extinction by dust in diffuse cirrus and in dense $``$birth clouds", following \citet{cha00}, is also applied (for details see SGPD11). SGPD11 found, in agreement with other studies, that they had to adopt dust parameters that varied with redshift in order to match the UV and B-band luminosity functions at high redshift. We note that unlike simple pure luminosity evolution models, SAMs have many physical sources of scatter in galaxy number densities and properties.

We compare our number counts to the SGPD11 model for several reasons. This model includes what are believed to be the key physical processes that shape galaxy formation and evolution. In particular, the FUV number counts are expected to provide an important constraint on the processes that trigger and regulate star formation, which are highly uncertain. The FUV number counts are also highly sensitive to dust extinction, which is another uncertain ingredient in the SAMs.

Our measured FUV number counts are broadly consistent with the SGPD11 and the SB4/Ly$\alpha$-flat SED model over all magnitudes. As seen in Figure \ref{fuvnc}, the SB4/Ly$\alpha$-flat SED model appears lower than the SGPD11 SAM up to FUV$_{AB}\sim$26.5 after which the trend is reversed and the SGPD11 model is lower. The differences in the bright end of the models is most likely due to the fact that the SB4/Ly$\alpha$-flat SED model is derived from a single spectral-energy distribution. Both sets of GALEX counts are lower than the SGPD11 model at the bright end, however the HAM10 field counts start to coincide with the models at FUV$_{AB}$ $>$ 22.5. This is consistent with the fact that the SGPD11 model is known to overproduce bright galaxies compared to GALEX data \citep{gil09,som11}, due to a small degree of residual $``$overcooling" in massive halos. Our number counts do not match the SB4/Ly$\alpha$-flat SED model at all magnitudes, but begin to coincide with it after 24.5 mag. As discussed by \citet{tep06}, the discrepancies with this model, especially towards bright magnitudes, may suggest a need for number density evolution in FUV galaxy number count models because this model only takes into account luminosity evolution.

\subsection{Effects of Cosmic Variance}
\label{cosmicvar}
\begin{figure*}
\epsscale{1}
\plotone{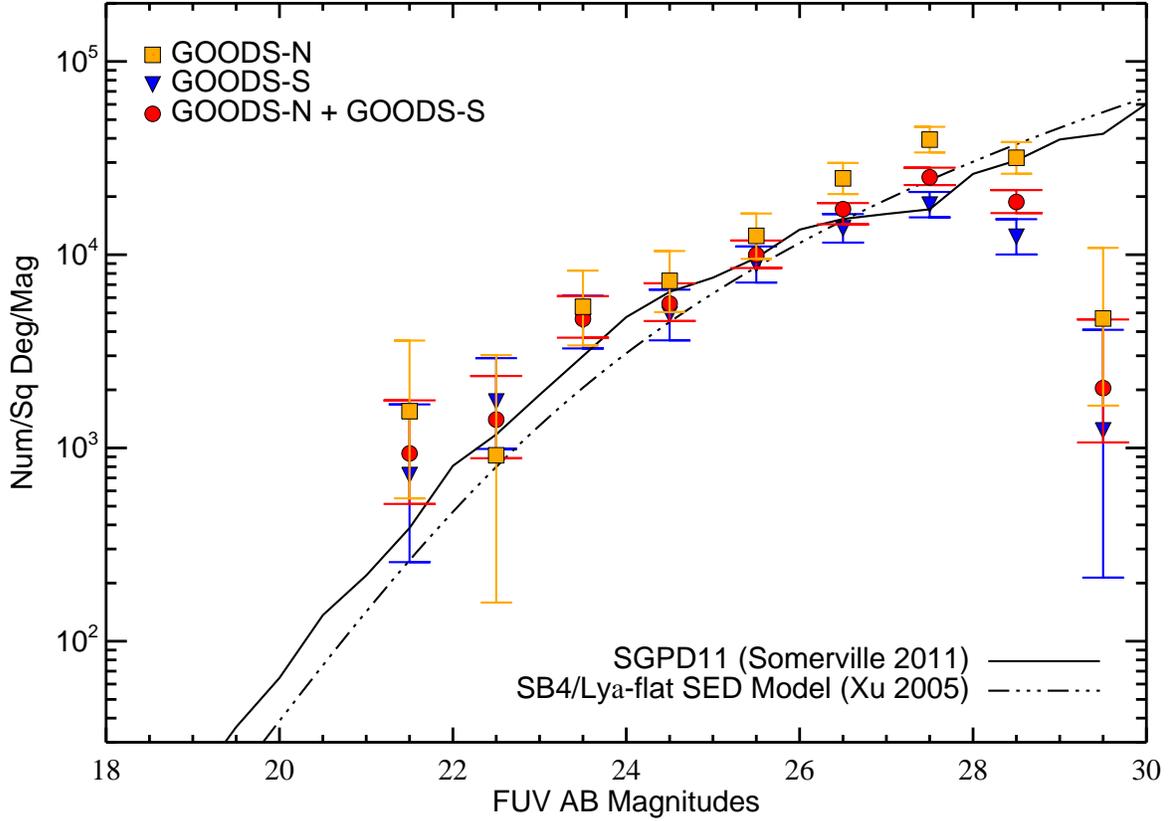}
\caption{FUV number counts for individual fields. We excluded the brightest (21.5 mag) and the faintest (29.5 mag) magnitude bins from this plot because there is not
enough signal-to-noise to make a comparison between fields at these magnitudes. HDF-N counts are from \citet{tep06}. The caps of the error bars do not reflect an error in magnitude, but have been manually varied in length to better distinguish amongst them.}
\label{cosvar}
\end{figure*}

Uncertainties in measurements of galaxy number counts can arise as a result of overall large-scale structure variation, or cosmic variance \citep{som04}.  The observations used for this study were designed to significantly reduce the effects of cosmic variance by including data from various sight-lines and covering a larger area than any previous FUV number counts study at these wavelengths and magnitudes. Our observations cover a total area of 15.9 arcmin$^{2}$, while the \citet{gar00} STIS observations  in the HDF-N and -S cover only 1.54 arcmin$^{2}$ and the \citet{tep06} SBC observations in the HDF-N cover only $\sim$3.77 arcmin$^{2}$. Also, the HDF-N has galaxy overdensities at z $\sim$0.45 and z $\sim$0.8 \citep{coh00} that bias the number counts in that field. To demonstrate the effects of cosmic variance we have compared in Figure \ref{cosvar} our total FUV number counts with the number counts calculated in the GOODS-N and GOODS-S SBC fields separately. The red circles represent the total number counts, blue upside down triangles represent the number counts in the GOODS-S area, and the orange squares represent the number counts in the GOODS-N area. The counts in the GOODS-N area are consistently higher than those in GOODS-S in every magnitude, bin except 22.5. The total number counts are a clear average of the number counts in these two fields over the entire magnitude range. This result demonstrates that using large areas and various sight-lines to make measurements of number counts reduces bias due to cosmic variance, and ideally these types of data sets provide the best comparisons for SAMs.

\subsection{The FUV Background Light from Resolved Sources}
\label{ebl}
The total UV background light is composed of several ingredients, broadly including emissions from the Earth's atmosphere, or airglow,  Galactic emissions, and extragalactic emissions. The Galactic component has been shown to be dominated by interstellar UV radiation scattered isotropically by dust, but also includes molecular hydrogen fluorescence, HII two-photon emission, and hot gas line emission, in smaller quantities \citep{mur09,bow91}. The extragalactic component is dominated by UV flux from resolved sources (i.e. galaxies), but may also include weak emission from the intergalactic medium (IGM). Measurements of the resolved UV extragalactic background light (EBL) can be determined from catalogs of extragalactic sources, and can be interpreted as an average measurement of the star-formation rate density over cosmological time, setting a lower limit for the total UV background light. Commonly, measurements of the UV background radiation that do not directly include these resolved sources are termed `diffuse background' measurements. Earlier studies making measurements of the diffuse FUV background are discussed in thorough reviews by \citet{bow91} and \citet{hen91}, while more recent work has been reviewed by \citet{mur09}. The definition of FUV wavelength coverage for each study varies between 912--1740\AA, depending on the detector used. 

Several techniques have been imparted in order to measure the diffuse FUV background. First, many studies have measured Galactic dust scattering, removing airglow effects, and fitting models to diffuse observations, extrapolating the signal down to zero column density (N$_{HI}$=0) which provides levels for what is interpreted as the FUV extragalactic background (i.e. galactic sources and potentially diffuse IGM emission). \citet{hen93} used this technique to reanalyze data from the Johns Hopkins UVX experiment for observations above $|b|=40\,^{\circ}$ (where b is Galactic latitude). An improved model simulating scattering of diffuse galactic light in the ISM was developed and used by \citet{wit94} to re-measure the extragalactic background in Dynamic Explorer 1 observations from \citet{fix89}. This same model was used by \citet{wit97} to re-evaluate the extragalactic background extrapolation from Far-Ultraviolet Space Telescope (FAUST) observations \citep{sas95}. \citet{sch01} derived the extragalactic FUV background with data from the Narrowband Ultraviolet Imaging Experiment (NUVIEWS), the first experiment primarily designed to map the FUV background. Most recently, this extrapolation technique has been used by \citet{seo10} to measure the FUV extragalactic background with the Spectroscopy of Plasma Evolution from Astrophysical Radiation instrument (SPEAR/FIMS). A second technique, that measures the truly diffuse extragalactic background, has been imparted by \citet{bro00} who masked the resolved FUV sources down to m$_{AB}$ = 29 in HST STIS HDF-N and -S, and HDF-N parallel imaging \citep{gar00}. They found a large unresolved diffuse background component that may include contributions from airglow. 

Other studies have used FUV spectra and imaging from large data sets to map the FUV background over a large range of Galactic latitudes, revealing patchy skymaps of the background due to variations in intensities of the flux at different latitudes. \citet{mur99} mapped the FUV background over the sky from 17 years of Voyager observations with the Voyager Ultraviolet Spectrometer (UVS), unique in that they are not partial to airglow effects, and \citet{mur04} mapped the FUV background intensity with Far-Ultraviolet Spectroscopic Explorere (FUSE) observations in 71 independent fields. Most recently, \citet{mur10} used archival GALEX imaging to map the diffuse FUV background over $\sim$75\% of the sky. This technique is used to put an upper limit on the extragalactic FUV background from values determined in the darkest areas of these data sets, primarily, but not necessarily, found in the vicinity of the Galactic poles. These FUV background skymaps have also revealed that some of the brightest FUV intensities are correlated with Galactic structures such as molecular clouds and nebulae. Detailed analysis to disentangle components of and effects on the diffuse FUV background in the vicinity of these structures have been carried out by determining correlations with HI column, H$_2$ fluorescence, Galactic extinction, and dust scattering, in some cases, resulting in measurments of a FUV extragalactic background component \citep{suj07,lee06,suj05}. Measurements of the diffuse FUV background are complimented by measurements of the resolved FUV background from extragalactic sources.    

\begin{figure*}
\epsscale{1}
\plotone{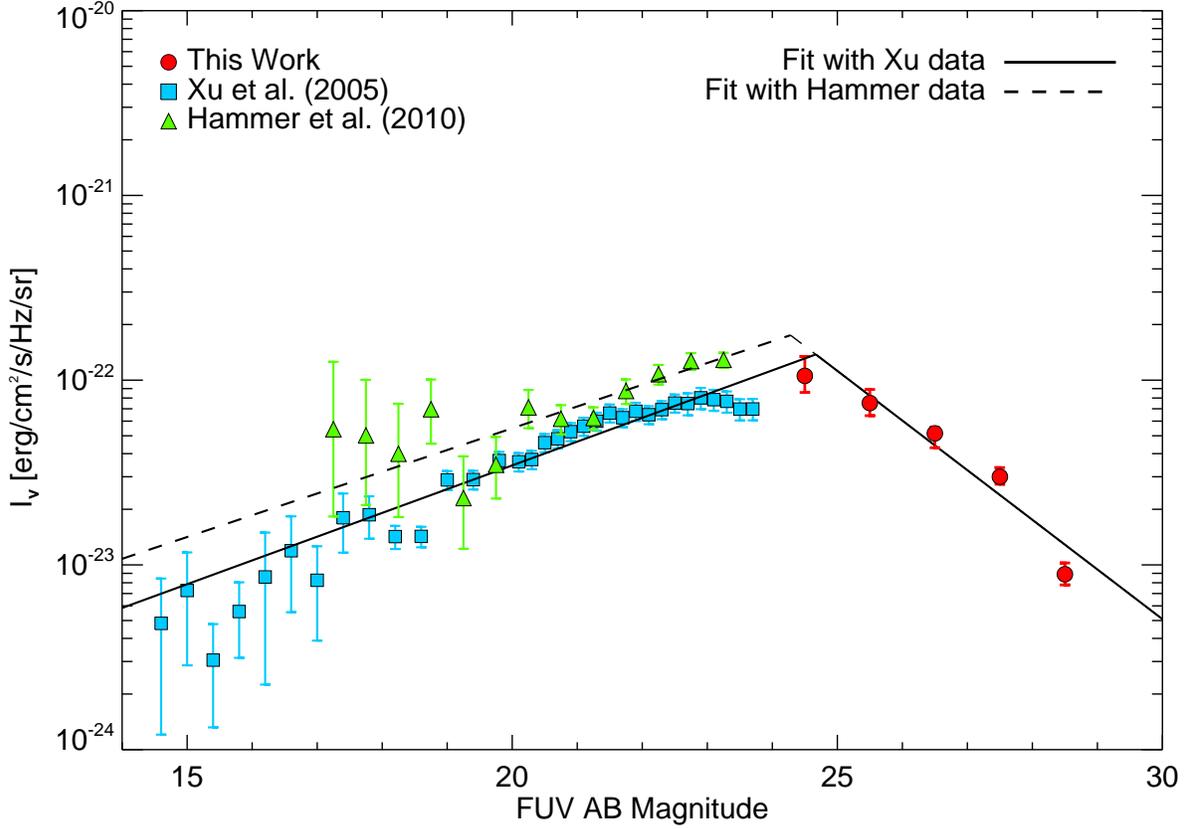}
\caption{Extragalactic background light from resolved sources per magnitude as a function of FUV magnitude. Two measurements are made from these data. The solid line measures the integrated EBL using the Xu et al. (2005) counts for the bright end (EBL I), while the dashed line makes this measurement using the Hammer et al. (2010) counts at the bright end (EBL II).}
\label{eblpmag}
\end{figure*}

In this study we calculate the FUV extragalactic background light (EBL) from resolved sources in the FUV data used for number counts, and these results are given in Table \ref{ebltab}. We use both sets of bright GALEX number counts in our calculation, giving us two possible values for the integrated EBL. First, we fit a slope of 0.13 $\pm$ 0.05 with an intercept of 0.68 $\pm$ 1.23 to our FUV number counts for magnitudes 24.5--28.5, including only the faint-end of the SBC/FUV number counts distribution. We also fit a slope of 0.53 $\pm$ 0.01 with an intercept of -9.11 $\pm$ 0.28 to the XU05 GALEX counts for magnitudes 14.2--23.7. For the HAM10 field GALEX counts we use the slope of 0.5 fitted to the FUV data by \citet{ham10} with an intercept of -8.7 $\pm$ 0.81 for magnitudes 17.25--23.25. Next, these slopes, as well as the number counts, are converted to units of EBL per magnitude bin, erg s$^{-1}$ cm$^{-2}$ Hz$^{-1}$ sr$^{-1}$, using the formula from \citet{mad00}:
\begin{equation}
I_{\nu} = 10^{-0.4(FUV_{AB} + 48.6)}N(FUV_{AB})
\end{equation}
Finally, we integrate under each function. For the combined fit of the faint-end SBC/FUV data with the XU05 data (hereafter EBL I), we set FUV$_{AB}$ = 24.67 as the upper limit for the integral of the XU05 function and the lower limit for the integral of the SBC/FUV function, because this magnitude is the maximum in integrated light. From this model we measure the integrated EBL for the magnitude range FUV$_{AB}$ = 14--30 of $\nu\mathit{I_{\nu}}$ = 1.3$^{+0.2}_{-0.2}$ nW m$^{-2}$ sr$^{-1}$, or in photon units, $\mathit{I_{\lambda}}$ = 65.9$^{+8}_{-8}$ photons s$^{-1}$ cm$^{-2}$ sr$^{-1}$ \AA $^{-1}$. The errors are 1 sigma uncertainties on the number counts. The models and data are plotted in Figure \ref{eblpmag}. The model for the GALEX data between FUV$_{AB}$=14-24.67 accounts for 66.5\% of EBL I, measuring more of the resolved background light than our faint-end number counts. For the combined fit of the faint-end SBC/FUV data with the HAM10 data (hereafter EBL II), we set FUV$_{AB}$ = 24.28 as the upper limit for the integral of the HAM10 function and the lower limit for the integral of the SBC/FUV function. From this model we measure the integrated EBL for the magnitude range FUV$_{AB}$ = 17--30 of $\nu\mathit{I_{\nu}}$ = 1.6$^{+0.2}_{-0.2}$ nW m$^{-2}$ sr$^{-1}$, or in photon units, $\mathit{I_{\lambda}}$ = 82.6$^{+12}_{-12}$ photons s$^{-1}$ cm$^{-2}$ sr$^{-1}$ \AA $^{-1}$.  Again, the GALEX portion of the model measures more resolved background light than our number counts, accounting for 66\% of EBL II, very similar to XU05. This similarity is due to a caveat in the data included from these two studies in that XU05 covers a larger magnitude range than HAM10, and the latter has higher I$_{\nu}$. This can be clearly seen in Figure \ref{eblpmag}. 

One of the first attempts at determining the FUV background from light emitted by galaxies was carried out by \citet{mar89}. They obtained data from an FUV imaging experiment  that used a rocket mounted detector to observe signatures of galaxies in the integrated FUV background. The experiment covered wavelengths 1350-1900\AA\ and determined a 1 sigma upper limit for the summed FUV intensity coming from sources $\sim$50 photons s$^{-1}$ cm$^{-2}$ sr$^{-1}$ \AA $^{-1}$, that is $\sim$25--40\% lower than our measurements. The UV EBL was measured at 2000\AA\ by \citet{mil92} from FOCA number counts and by \citet{arm94} from predictions of number counts \citep{2arm94}. While our measurements are well within the range of 40--130 photons s$^{-1}$ cm$^{-2}$ sr$^{-1}$ \AA $^{-1}$ predicted by \citet{arm94}, they are much higher than the 23 photons s$^{-1}$ cm$^{-2}$ sr$^{-1}$ \AA $^{-1}$ determined from the FOCA number counts between magnitudes 15.0--18.5. Comparing our measurements to those from \citet{gar00}, EBL I and EBL II are $\sim$54--66$\%$ and $\sim$43--58$\%$ lower, respectively, than their measurements of 2.9$_{-0.4}^{+0.6}$--3.9$_{-0.8}^{+1.1}$ nW m$^{-2}$ sr$^{-1}$ (144$_{-19}^{+28}$--195$_{-39}^{+59}$ photons s$^{-1}$ cm$^{-2}$ sr$^{-1}$ \AA $^{-1}$) at 1595\AA. \citet{xu05} extrapolated models fit to the GALEX FUV number counts (1530\AA), integrated these functions to zero flux, and measured the total FUV EBL to be 1.03 $\pm$ 0.15 nW m$^{-2}$ sr$^{-1}$ which is $\sim$21$\%$ lower than EBL I, $\sim$37$\%$ lower than EBL II, and also below the \citet{gar00} range. The conclusion that can be drawn from our measurments is that the resolved EBL is unlikely to be much greater than $\sim$100 photons s$^{-1}$ cm$^{-2}$ sr$^{-1}$ \AA $^{-1}$, and therefore other diffuse EBL measurements with significantly higher values \citep{sch01,bro00,wit97,wit94,hen93} almost certainly include Galactic contributions and potentially smaller contributions from airglow. All values for the resolved FUV EBL discussed in this section are summarized in Table \ref{ebltab}.

\section{Summary}
\label{sum}
We have presented FUV galaxy number counts at 1614\AA\ measured from deep HST ACS/SBC observations of the HUDF area of the GOODS-S field, the HDF-N area of the GOODS-N field, and 15 smaller fields at various pointings in GOODS-N and -S. We sample the faint-end of the FUV number counts out to FUV$_{AB}$ $\sim$29, with the majority of the sources in magnitude bins 23.5-28.5, and cover an area (15.9 arcmin$^{2}$) $\sim$4 times larger than the most recent deep FUV number counts survey \citep{tep06} at these wavelength and magnitude ranges. 
The number counts distribution provides the following results:

1. A slope of 0.13 $\pm$ 0.04 (intercept of 0.68 $\pm$ 1.23) fits the faint-end of the logarithmic number counts distribution from FUV$_{AB}$ = 24.5 to 28.5.

2. These number counts are $\sim$35\% and $\sim$36\% lower, on average, than the faint FUV counts measured in the HDF-N area of GOODS-N from \citet{gar00} and \citet{tep06}, respectively. The differences are most likely due to cosmic variance.

3. The bright end of the number counts slope, at FUV$_{AB}$ = 22.5, intersects the most recent GALEX FUV number counts from \citet{ham10}, but is higher than the GALEX FUV counts from \citet{xu05} at all common magnitudes. 

4. The latest $\lambda$CDM semi-analytic model based on the WMAP5 cosmology \citep{som11} is in good agreement with the FUV number counts. Generally, the FUV counts are higher than the SB4/Ly$\alpha$-flat single SED model \citep{xu05} but become more consistent at the faint-end. This may result from the model being based on a single starburst SED, thus offering evidence for number density evolution.

5. The integrated light from field galaxies contributes 1.3$^{+0.2}_{-0.2}$ nW m$^{-2}$ sr$^{-1}$ or 65.9$^{+8}_{-8}$ photons s$^{-1}$ cm$^{-2}$ sr$^{-1}$ \AA $^{-1}$ to the FUV extragalactic background light for magnitudes 14-30 when measured with XU05 bright-end GALEX counts, and 1.6$^{+0.2}_{-0.2}$ nW m$^{-2}$ sr$^{-1}$ or 82.6$^{+12}_{-12}$ photons s$^{-1}$ cm$^{-2}$ sr$^{-1}$ \AA $^{-1}$ for magnitudes 17-30 when measured with HAM10 bright-end GALEX counts. The GALEX portion of these models account for $\sim$66\% of the total integrated light in each case. This measurement sets a lower limit for future calculations of the diffuse background. The resolved EBL is unlikely to be much greater than $\sim$100 photons s$^{-1}$ cm$^{-2}$ sr$^{-1}$ \AA $^{-1}$. Any measurement that yields values significantly higher than this value almost certainly includes Galactic and airglow contributions.



\acknowledgments

We are grateful to the anonymous referee for their helpful comments that improved this paper. We would like to thank R.S. Somerville and R.C. Gilmore for providing us with their semi-analytic model and for numerous helpul discussions and comments on this paper. We would also like to thank D. Hammer for useful science discussions and sharing number counts results, and C.K. Xu for providing us with theoretical number counts models. Support for Program numbers GO-10403 and GO-10872 was provided by NASA through a grant from the Space Telescope Science Institute, which is operated by the Association of Universities for Research in Astronomy, Incorporated, under NASA contract NAS5-26555. E.N.V. was funded by the NASA Graduate Student Research Program grant \#NNX08AR95H.

This research has made use of data obtained from the Chandra Source Catalog, provided by the Chandra X-ray Center (CXC) as part of the Chandra Data Archive.

This research has made use of the NASA/IPAC Infrared Science Archive, which is operated by the Jet Propulsion Laboratory, California Institute of Technology, under contract with the National Aeronautics and Space Administration.

\clearpage

\clearpage
\begin{deluxetable}{cccccccc}
\tabletypesize{\small}
\tablewidth{0.0pt}
\tablenum{1}
\tablecaption{\sc{FUV Galaxy Counts}}
\tablecolumns{8}
\tablehead{
       \colhead{FUV$_{AB}$} &
       \colhead{NC} &
       \colhead{\null} &
       \colhead{\null} &
       \colhead{\null} &
       \colhead{\null} &
       \colhead{\null}&
       \colhead{Area}\\
  \colhead{(mag)} &
  \colhead{(No. deg$^{-2}$ mag$^{-1}$)} &
  \colhead{log NC} &
  \colhead{$\sigma_{low}$} &
  \colhead{$\sigma_{high}$} &
  \colhead{Raw No.} &
  \colhead{Completeness}&
  \colhead{(arcmin$^{2}$)}
}

\startdata
21.5&937&2.97&0.26&0.27&4&1.054&13.89\\
22.5&1402&3.15&0.20&0.23& 6&1.056&13.89\\
23.5&4656&3.67&0.10&0.12&20&1.026&14.70\\
24.5&5582&3.75&0.09&0.11&24&1.019&14.91\\
25.5&9996&4.00&0.07&0.07&43&1.007&15.28\\
26.5&17207&4.24&0.08&0.03&74&1.003&15.42\\
27.5&25166&4.40&0.04&0.05&99&0.972&15.20\\
28.5&18752&4.27&0.06&0.06&60&0.854&14.33\\
29.5&2041&3.31&0.28&0.36&3&0.317&14.56\\
\enddata
\tablenotetext{\null}{Note---Magnitudes represent the center of the bins, errors are 1$\sigma$ Poisonnian \citep{geh86}, and areas are the average total detection areas of all objects within each magnitude bin.} 
\label{nctab}
\end{deluxetable}

\clearpage
\begin{deluxetable}{cccccc}
\tabletypesize{\scriptsize}
\tablewidth{0.0pt}
\tablenum{2}
\tablecaption{\sc{Measurements of the Resolved FUV Background Light}}
\tablecolumns{6}
\tablehead{
       \colhead{Investigators} &
       \colhead{Instrument} &
       \colhead{$\lambda$} &
       \colhead{Magnitudes Covered}&
       \colhead{FUV BL} &
       \colhead{FUV BL}  \\
  \colhead{\null} &
  \colhead{\null} &
  \colhead{(\AA)} &
   \colhead{(AB)}&
  \colhead{(nW/m$^{2}$/sr)}&
  \colhead{(phot/s/cm$^{2}$/sr/\AA)}
}

\startdata
This Work: EBL I\tablenotemark{a}&SBC/GALEX&1614/1530&14.70--29.30&1.3$^{+0.2}_{-0.2}$&65.9$^{+8}_{-8}$\\
This Work: EBL II\tablenotemark{b}&SBC/GALEX&1614/1530&17.30--29.70&1.6$^{+0.2}_{-0.2}$&82.6$^{+12}_{-12}$\\
\citet{xu05}&GALEX&1530&extrap. to zero mag&1.03 $\pm$ 0.15&52 $\pm$ 7\\
\citet{gar00}&STIS/FOCA&1595&17.50--29.50&2.9$_{-0.4}^{+0.6}$ to 3.9$_{-0.8}^{+1.1}$&144$_{-19}^{+28}$ to 195$_{-39}^{+59}$\\
\citet{arm94}&...\tablenotemark{c}&2000&15.00--18.50&0.8 to 2.6&40 to 130\\
\citet{mil92}&FOCA&2000&15.00--18.50&0.4&23\\
\enddata
\tablenotetext{a}{Bright end fit is from \citet{xu05} GALEX FUV number counts.}
\tablenotetext{b}{Bright end fit is from \citet{ham10} GALEX FUV number counts.}
\tablenotetext{c}{his measurement is from a
prediction of number counts based on galaxy
evolution models and published galaxy SEDs.}
\label{ebltab}
\end{deluxetable}


\begin{thebibliography}{}
\bibitem[Armand \& Milliard(1994)]{2arm94} Armand, C., \& Milliard, B.\ 1994, \aap, 282, 1
\bibitem[Armand et al.(1994)]{arm94} Armand, C., Milliard, B., \& Deharveng, J.~M.\ 1994, \aap, 284, 12
\bibitem[Arnouts et al.(2005)]{arn05} Arnouts, S., et al.\ 2005, \apjl, 619, L43
\bibitem[Beckwith et al.(2006)]{bec06} Beckwith, S.~V.~W., et al.\ 2006, \aj, 132, 1729
\bibitem[Bertin \& Arnouts(1996)]{ber96} Bertin, E., \& Arnouts, S.\ 1996, \aaps, 117, 393
\bibitem[Bowyer(1991)]{bow91} Bowyer, S.\ 1991, \araa, 29, 59
\bibitem[Bridge et al.(2010)]{bri10} Bridge, C.~R., et al.\ 2010, \apj, 720, 465
\bibitem[Brown et al.(2000)]{bro00} Brown, T.~M., Kimble, R.~A., Ferguson, H.~C., Gardner, J.~P., Collins, N.~R., \& Hill, R.~S.\ 2000, \aj, 120, 1153
\bibitem[Bruzual \& Charlot(2003)]{bru03} Bruzual, G., \& Charlot, S.\ 2003, \mnras, 344, 1000
\bibitem[Cardelli et al.(1989)]{car89} Cardelli, J.~A., Clayton, G.~C., \& Mathis, J.~S.\ 1989, \apj, 345, 245
\bibitem[Charlot \& Fall(2000)]{cha00} Charlot, S., \& Fall, S.~M.\ 2000, \apj, 539, 718
\bibitem[Cohen et al.(2000)]{coh00} Cohen, J.~G., Hogg, D.~W., Blandford, R., Cowie, L.~L., Hu, E., Songaila, A., Shopbell, P., \& Richberg, K.\ 2000, \apj, 538, 29
\bibitem[Cohen(2001)]{coh01} Cohen, J.~G.\ 2001, \aj, 121, 2895
\bibitem[Cowie et al.(2009)]{cow09} Cowie, L.~L., Barger, A.~J., \& Trouille, L.\ 2009, \apj, 692, 1476
\bibitem[Dahlen et al.(2007)]{dah07} Dahlen, T., Mobasher, B., Dickinson, M., Ferguson, H.~C., Giavalisco, M., Kretchmer, C., \& Ravindranath, S.\ 2007, \apj, 654, 172
\bibitem[Dahlen et al.(2010)]{dah10} Dahlen, T., et al.\ 2010, \apj, 724, 425
\bibitem[Dawson et al.(2001)]{daw01} Dawson, S., Stern, D., Bunker, A.~J., Spinrad, H., \& Dey, A.\ 2001, \aj, 122, 598
\bibitem[de Mello et al.(2004)]{dem04} de Mello, D.~F., Gardner, J.~P., Dahlen, T., Conselice, C.~J., Grogin, N.~A., \& Koekemoer, A.~M.\ 2004, \apjl, 600, L151
\bibitem[Deharveng et al.(1994)]{deh94} Deharveng, J.-M., Sasseen, T.~P., Buat, V., Bowyer, S., Lampton, M., \& Wu, X.\ 1994, \aap, 289, 715
\bibitem[Deharveng et al.(2001)]{deh01} Deharveng, J.-M., Buat, V., Le Brun, V., Milliard, B., Kunth, D., Shull, J.~M., \& Gry, C.\ 2001, \aap, 375, 805
\bibitem[Evans et al.(2010)]{eva10} Evans, I.~N., et al.\ 2010, \apjs, 189, 37
\bibitem[Fan et al.(2006)]{fan06} Fan, X., et al.\ 2006, \aj, 132, 117
\bibitem[Fern{\'a}ndez-Soto et al.(2003)]{fer03} Fern{\'a}ndez-Soto, A., Lanzetta, K.~M., \& Chen, H.-W.\ 2003, \mnras, 342, 1215
\bibitem[Fix et al.(1989)]{fix89} Fix, J.~D., Craven, J.~D., \& Frank, L.~A.\ 1989, \apj, 345, 203
\bibitem[Gardner et al.(2000a)]{gar00a} Gardner, J.~P., et al.\ 2000, \aj, 119, 48
\bibitem[Gardner et al.(2000b)]{gar00} Gardner, J.~P., Brown, T.~M., \& Ferguson, H.~C.\ 2000, \apjl, 542, L79
\bibitem[Gehrels(1986)]{geh86} Gehrels, N.\ 1986, \apj, 303, 336
\bibitem[Giavalisco et al.(2004)]{gia04} Giavalisco, M., et al.\ 2004, \apjl, 600, L93
\bibitem[Gilmore et al.(2009)]{gil09} Gilmore, R.~C., Madau, P., Primack, J.~R., Somerville, R.~S., \& Haardt, F.\ 2009, \mnras, 399, 1694
\bibitem[Hammer et al.(2010)]{ham10} Hammer, D., et al.\ 2010, \apjs, 191, 143
\bibitem[Henry(1991)]{hen91} Henry, R.~C.\ 1991, \araa, 29, 89
\bibitem[Henry \& Murthy(1993)]{hen93} Henry, R.~C., \& Murthy, J.\ 1993, \apjl, 418, L17
\bibitem[Hoversten et al.(2009)]{hov09} Hoversten, E.~A., et al.\ 2009, \apj, 705, 1462
\bibitem[Iglesias-P{\'a}ramo et al.(2004)]{igl04} Iglesias-P{\'a}ramo, J., Buat, V., Donas, J., Boselli, A., \& Milliard, B.\ 2004, \aap, 419, 109
\bibitem[Inoue et al.(2006)]{ino06} Inoue, A.~K., Iwata, I., \& Deharveng, J.-M.\ 2006, \mnras, 371, L1 
\bibitem[Iwata et al.(2009)]{iwa09} Iwata, I., et al.\ 2009, \apj, 692, 1287
\bibitem[Kinney et al.(1996)]{kin96} Kinney, A.~L., Calzetti, D., Bohlin, R.~C., McQuade, K., Storchi-Bergmann, T., \& Schmitt, H.~R.\ 1996, \apj, 467, 38
\bibitem[Lee et al.(2006)]{lee06} Lee, D.-H., et al.\ 2006, \apjl, 644, L181
\bibitem[Madau \& Pozzetti(2000)]{mad00} Madau, P., \& Pozzetti, L.\ 2000, \mnras, 312, L9
\bibitem[Martin \& Bowyer(1989)]{mar89} Martin, C., \& Bowyer, S.\ 1989, \apj, 338, 677
\bibitem[Milliard et al.(1992)]{mil92} Milliard, B., Donas, J., Laget, M., Armand, C., \& Vuillemin, A.\ 1992, \aap, 257, 24
\bibitem[Murthy et al.(1999)]{mur99} Murthy, J., Hall, D., Earl, M., Henry, R.~C., \& Holberg, J.~B.\ 1999, \apj, 522, 904
\bibitem[Murthy \& Sahnow(2004)]{mur04} Murthy, J., \& Sahnow, D.~J.\ 2004, \apj, 615, 315
\bibitem[Murthy(2009)]{mur09} Murthy, J.\ 2009, \apss, 320, 21
\bibitem[Murthy et al.(2010)]{mur10} Murthy, J., Henry, R.~C., \& Sujatha, N.~V. \ 2010, arXiv:1009.4530
\bibitem[Pimbblet et al.(2004)]{pim04} Pimbblet, K.~A., Drinkwater, M.~J., \& Hawkrigg, M.~C.\ 2004, \mnras, 354, L61
\bibitem[Schiminovich et al.(2001)]{sch01} Schiminovich, D., Friedman, P.~G., Martin, C., \& Morrissey, P.~F.\ 2001, \apjl, 563, L161
\bibitem[Sasseen et al.(1995)]{sas95} Sasseen, T.~P., Lampton, M., Bowyer, S., \& Wu, X.\ 1995, \apj, 447, 630
\bibitem[Seon et al.(2010)]{seo10} Seon, K.~-., et al.\ 2010, arXiv:1006.4419
\bibitem[Shapley et al.(2006)]{sha06} Shapley, A.~E., Steidel, C.~C., Pettini, M., Adelberger, K.~L., \& Erb, D.~K.\ 2006, \apj, 651, 688
\bibitem[Schlegel et al.(1998)]{shl98} Schlegel, D.~J., Finkbeiner, D.~P., \& Davis, M.\ 1998, \apj, 500, 525
\bibitem[Siana et al.(2007)]{sia07} Siana, B., et al.\ 2007, \apj, 668, 62
\bibitem[Siana et al.(2010)]{sia10} Siana, B., et al.\ 2010, arXiv:1001.3412 
\bibitem[Somerville \& Primack(1999)]{som99} Somerville, R.~S., \& Primack, J.~R.\ 1999, \mnras, 310, 1087
\bibitem[Somerville et al.(2001)]{som01} Somerville, R.~S., Primack, J.~R., \& Faber, S.~M.\ 2001, \mnras, 320, 504
\bibitem[Somerville et al.(2004)]{som04} Somerville, R.~S., Lee, K., Ferguson, H.~C., Gardner, J.~P., Moustakas, L.~A., \& Giavalisco, M.\ 2004, \apjl, 600, L171 
\bibitem[Somerville et al.(2008)]{som08} Somerville, R.~S., Hopkins, P.~F., Cox, T.~J., Robertson, B.~E., \& Hernquist, L.\ 2008, \mnras, 391, 481
\bibitem[Somerville et al.(2011)]{som11} Somerville R. S., Gilmore R. C., Primack J. R., Dominguez A., 2010, \emph{in preparation}
\bibitem[Steidel et al.(2001)]{ste01} Steidel, C.~C., Pettini, M., \& Adelberger, K.~L.\ 2001, \apj, 546, 665
\bibitem[Sujatha et al.(2005)]{suj05} Sujatha, N.~V., Shalima, P., Murthy, J., \& Henry, R.~C.\ 2005, \apj, 633, 257
\bibitem[Sujatha et al.(2007)]{suj07} Sujatha, N.~V., Murthy, J., Shalima, P., \& Henry, R.~C.\ 2007, \apj, 665, 363
\bibitem[Teplitz et al.(2006)]{tep06} Teplitz, H.~I., et al.\ 2006, \aj, 132, 853
\bibitem[Teplitz et al.(2010)]{tep10} Teplitz, H.~I., et al.\ 2010, \emph{in preparation}
\bibitem[Williams et al.(1996)]{wil96} Williams, R.~E., et al.\ 1996, \aj, 112, 1335
\bibitem[Witt \& Petersohn(1994)]{wit94} Witt, A.~N., \& Petersohn, J.~K.\ 1994, The First Symposium on the Infrared Cirrus and Diffuse Interstellar Clouds, 58, 91
\bibitem[Witt et al.(1997)]{wit97} Witt, A.~N., Friedmann, B.~C., \& Sasseen, T.~P.\ 1997, \apj, 481, 809
\bibitem[Wyder et al.(2005)]{wyd05} Wyder, T.~K., et al.\ 2005, \apjl, 619, L15
\bibitem[Xu et al.(2005)]{xu05} Xu, C.~K., et al.\ 2005, \apjl, 619, L11
\end{thebibliography}
\end{document}